# The Fluorescence Telescope on board EUSO-SPB2 for the detection of Ultra High Energy Cosmic Rays

**G. Osteria[a,*] J. Adams[b], M. Battisti[c], A. Belov[d], M. Bertaina[e,c], F. Bisconti[c], F. Cafagna[f], D. Campana[a], R. Caruso[gh], M. Casolino[ij], M. Christi[b], T. Ebisuzaki[j], J. Eser[k], F. Fenu[c], G. Filippatos[l], C. Fornaro[m], F. Guarino[n,a], P. Klimov[d], V. Kungel[l], S. Mackovjak[o] M. Mese[a], M. Miller[p], H. Miyamoto[c], A. Olinto[k], Y. Onel[p], E. Parizot[q], M. Pech[r], F. Perfetto[a], L. Piotrowski[s], G. Prevot[q], P. Reardon[b], M. Ricci[t], F. Sarazin[l], V. Scotti[n,a], K. Shinozaki[s], P. Shovanec[r], J. Szabelski[s], Y. Takizawa[j], L. Valore[n,a], L. Wiencke[l] on behalf of the JEM EUSO Collaboration**

(a complete list of authors can be found at the end of the proceedings)

[a]*Istituto Nazionale di Fisica Nucleare, Naples, Italy*

[b]*University of Alabama in Huntsville, Huntsville, AL, USA*

[c]*Istituto Nazionale di Fisica Nucleare, Turin, Italy*

[d]*Moscow State University, Moscow, Russia*

[e]*Università di Torino, Turin, Italy*

[f]*Istituto Nazionale di Fisica Nucleare, Bari, Italy*

[g]*Università di Catania, Catania, Italy*

[h]*Istituto Nazionale di Fisica Nucleare, Catania, Italy*

[i]*Istituto Nazionale di Fisica Nucleare, Roma Tor Vergata, Italy*

[j]*Riken, Wako, Japan*

[k]*University of Chicago, Chicago, IL, USA*

[l]*Colorado School of Mines, Golden, CO, USA*

[m]*Istituto Nazionale di Fisica Nucleare, Rome, Italy*

[n]*Università di Napoli Federico II, Naples, Italy*

[o]*Institute of Experimental Physics, Kosice, Slovakia*

[p]*University of Iowa, Iowa City, IA, USA*

[q]*APC, Univ Paris Diderot Paris, Paris Cité, France*

*Presenter

[r] *Palacký University Olomouc, Olomouc, Czech Republic*

[s]*National Centre for Nuclear Research, Lodz, Poland*

[t]*Istituto Nazionale di Fisica Nucleare, Frascati, Italy*

*E-mail:* giuseppe.osteria@na.infn.it,

The Fluorescence Telescope is one of the two telescopes on board the Extreme Universe Space Observatory on a Super Pressure Balloon II (EUSO-SPB2). EUSO-SPB2 is an ultra-long-duration balloon mission that aims at the detection of Ultra High Energy Cosmic Rays (UHECR) via the fluorescence technique (using a Fluorescence Telescope) and of Ultra High Energy (UHE) neutrinos via Cherenkov emission (using a Cherenkov Telescope). The mission is planned to fly in 2023 and is a precursor of the Probe of Extreme Multi-Messenger Astrophysics (POEMMA). The Fluorescence Telescope is a second generation instrument preceded by the telescopes flown on the EUSO-Balloon and EUSO-SPB1 missions. It features Schmidt optics and has a 1-meter diameter aperture. The focal surface of the telescope is equipped with a 6912-pixel Multi Anode Photo Multipliers (MAPMT) camera covering a 37.4 x 11.4 degree Field of Regard. Such a big Field of Regard, together with a flight target duration of up to 100 days, would allow, for the first time from suborbital altitudes, detection of UHECR fluorescence tracks. This contribution will provide an overview of the instrument including the current status of the telescope development.

## 1. Introduction

The EUSO-SPB2 *Fluorescence Telescope* (FT) is an instrument configured specifically for measuring the fluorescence component of Ultra High Energy Cosmic Rays (UHECRs) induced Extensive Air Showers (EAS). It is hosted on board the second-generation Extreme Universe Space Observatory on a Super-Pressure Balloon (EUSO-SPB2) mission [1-2]. The EUSO-SPB2 mission (planned to be launched in early 2023) is an important step toward the future space mission POEMMA (Probe Of Extreme Multi-Messenger Astrophysics) [3]. POEMMA will monitor the Earth's atmosphere to detect EASs produced by extremely energetic cosmic messengers: Ultra High Energy (UHE) neutrinos via Cherenkov emission, and UHECRs ($E > 10^{19}$ eV) via the fluorescence light. In order to allow the development and improvement of detection techniques for both signals (Cherenkov and fluorescence), EUSO-SPB2 will have on-board two telescopes (Fig. 1 left panel). The *Cherenkov telescope* (CT) [4] will measure air showers by imaging their Cherenkov light pointing near the Earth limb. The FT (Fig. 1 right panel), nadir pointing, will measure fluorescence light from EAS and is expected significantly contribute to the science goals of EUSO-SPB2 mission allowing for the first observation of extensive air showers using the fluorescence technique from suborbital space.

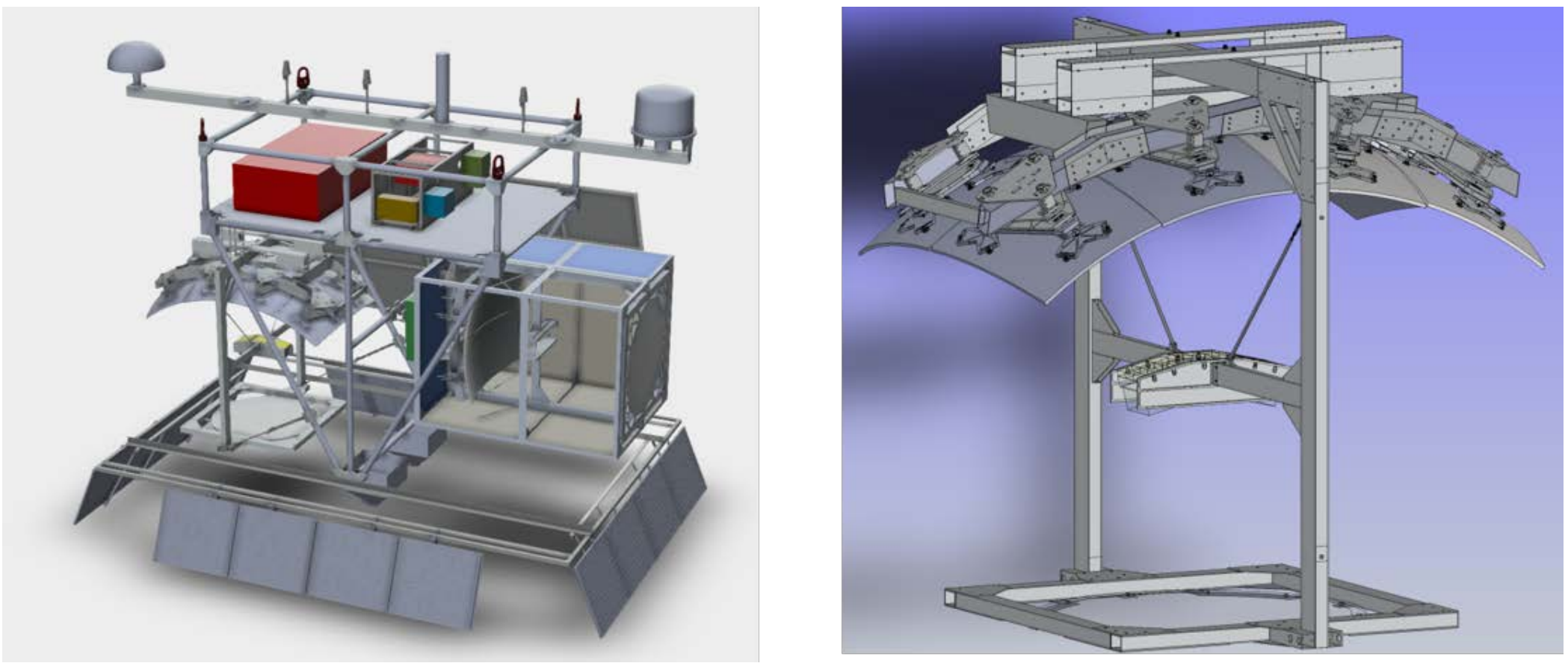

*Figure 1: Left: The EUSO SPB2 gondola with the two telescopes, the FT (nadir pointing) and CT (pointing at Earth limb). Right: The SPB2 Fluorescence Telescope (CAD rendering).*

## 2 The Fluorescence Telescope

The EUSO-SPB2 Fluorescence Telescope inherits from the experience of the previous missions developed by the JEM-EUSO Collaboration: EUSO-TA [5], EUSO-Balloon [6], and EUSO-SPB1 [7, 8]. With respect to the telescope which flew on the SPB1 mission, several upgrades on the optics, focal surface and general architecture have been introduced to improve the performance of the instrument. In particular, the focal surface is three times larger to increase the UHECR collection power. A more performing optics (Schmidt telescope) and a shorter temporal resolution of 1 μs have been implemented to decrease the energy threshold of the instrument.

Figure 2 shows a schematic view of the FT and its main subsystems. The hearth of the telescope is the Ultraviolet (UV) camera. It is mounted at the convex focal surface where the Schmidt optics focuses the light. It can count single photoelectrons in a wavelength bandwidth between 290 nm and 430 nm with an integration time of 1 μs and double pulse resolution of 6 ns.

To avoid damaging the focal plane sensors with direct sunlight, the telescope is equipped with a motorized cover that remains closed during the local day and is opened during the local night hours before the start of data acquisition operations.

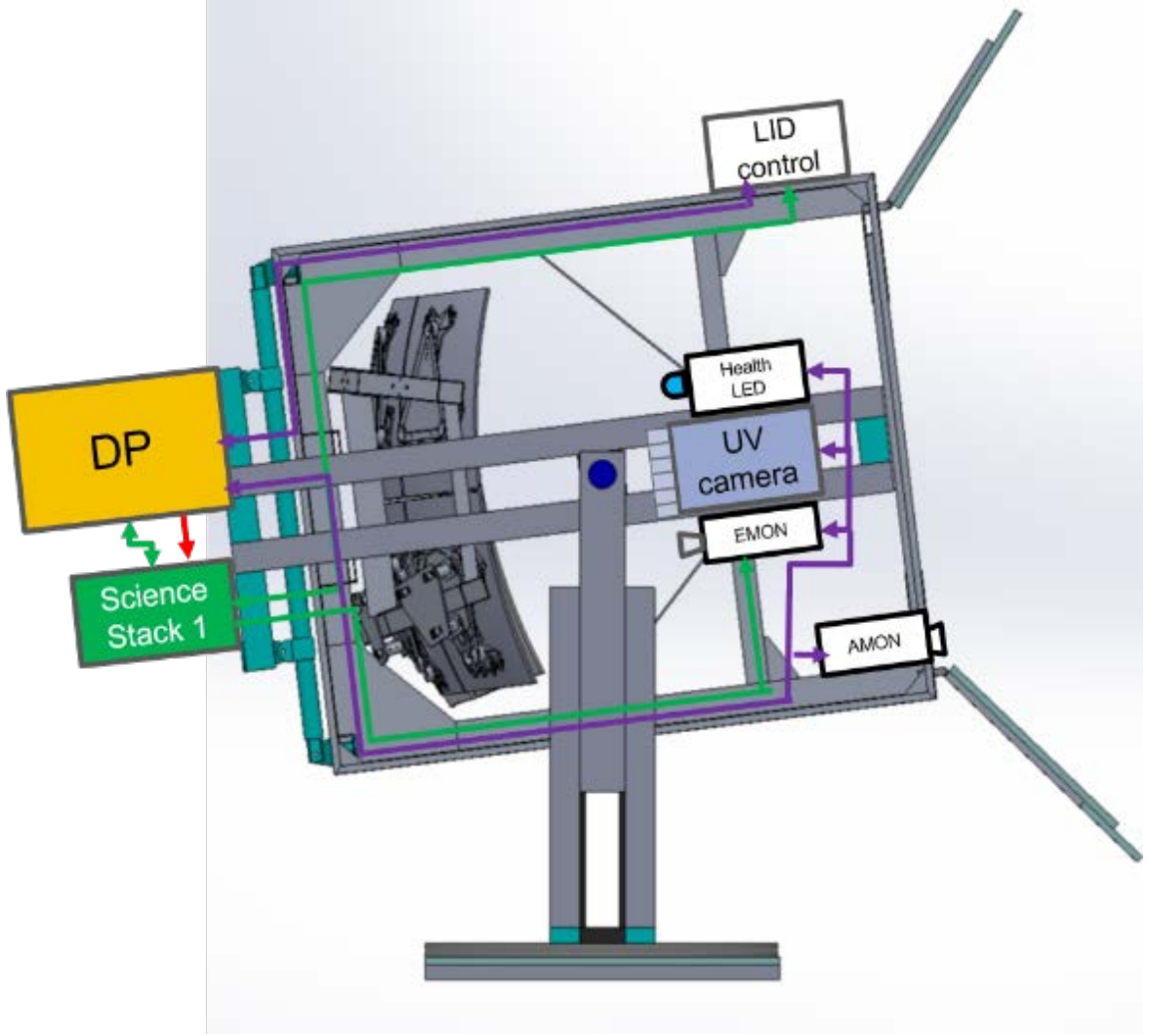

*Figure 2: FT main sub-systems*

The intensity of the light arriving on the telescope is measured by two independent devices. A first device (AMON), placed outside the telescope, is able to measure the absolute intensity of the UV / VIS radiation flow coming from the atmosphere and is also used to discriminate the day / night condition. The second device (EMON) is placed inside the telescope, near the focal surface and measures the intensity of the light arriving on it. It is used to check the dark condition inside the telescope before starting the measurement operations and to check the tightness of the lid closure. A system based on LEDs (Light Emission Diode) allows to illuminate the focal surface in a uniform manner to check the operating status of all the pixels of the focal plane sensors.

The Data Processor (DP) subsystem allows the remote control of all the sub-systems and manages the procedures for the acquisition and local storage of the data.

The details of the different sub-systems are described in the following sections.

## 2.1 The optics

The Schmidt optics has 6 mirrors segments with a radius of curvature of 1659.8 mm and an effective focal length of 860 mm. The aperture is 1 m diameter. It focuses the light in a 37.4°×11.4° field of regard at a convex focal surface on which the UV camera is placed. A plano-convex corrector lens will be placed in front of each PDM to preserve the point spread function (PSF).

## 2.2 The UV camera

The UV camera is the core of the instrument. It is a 6912-pixel camera segmented in three modules called Photo Detection Modules (PDMs). The three PDMs are mounted at the convex focal surface where the Schmidt optics focuses the light as shown in Figure 3.

Each PDM features 9 elementary cells (ECs) composed of four 64-pixel Multi Anodic Photo Multiplier Tubes (MAPMTs), covered with BG3 UV-band-pass filters, for a total of 2304 pixels. Each pixel is 3mm x 3mm wide.

The Elementary Cell is a compact assembly containing one HVPS generator board powering the four MAPMTs, the EC-anode boards and the ASICs for the signal digitation. Each EC is "potted" in a gelatinous compound to prevent discharge between the different components at the low pressure of EUSO-SPB2 altitude (Fig.4 left panel). The MAPMTs operate in photon counting mode. The anode pulse at the output of the MAPMT is processed by the corresponding ASIC (SPACIROC3 ASICs developed by Omega (CNRS, France)). The signal is discriminated, with a threshold settable for each MAPMT, and integrated in a 1.0μs period (defined to be 1 GTU). The SPACIROC3 ASIC allows a double pulse separation at 100 MHz with a power consumption of around 0.7 mW/channel. Every GTU, each ASIC sends the counts from the corresponding MAPMT to the UV camera electronics for readout.

### 2.2.1 The HVPS system.

The HVPS system, is composed of a HVPS control board and of one HVPS board per EC, which hosts the high voltage generators. The generators are Cockcroft-Walton voltage multipliers which provide the voltage to the dynodes and anodes of each MAPMTs of the EC. The HVPS board provides insulation and communication between the HVPS generators and the rest of the instrument.

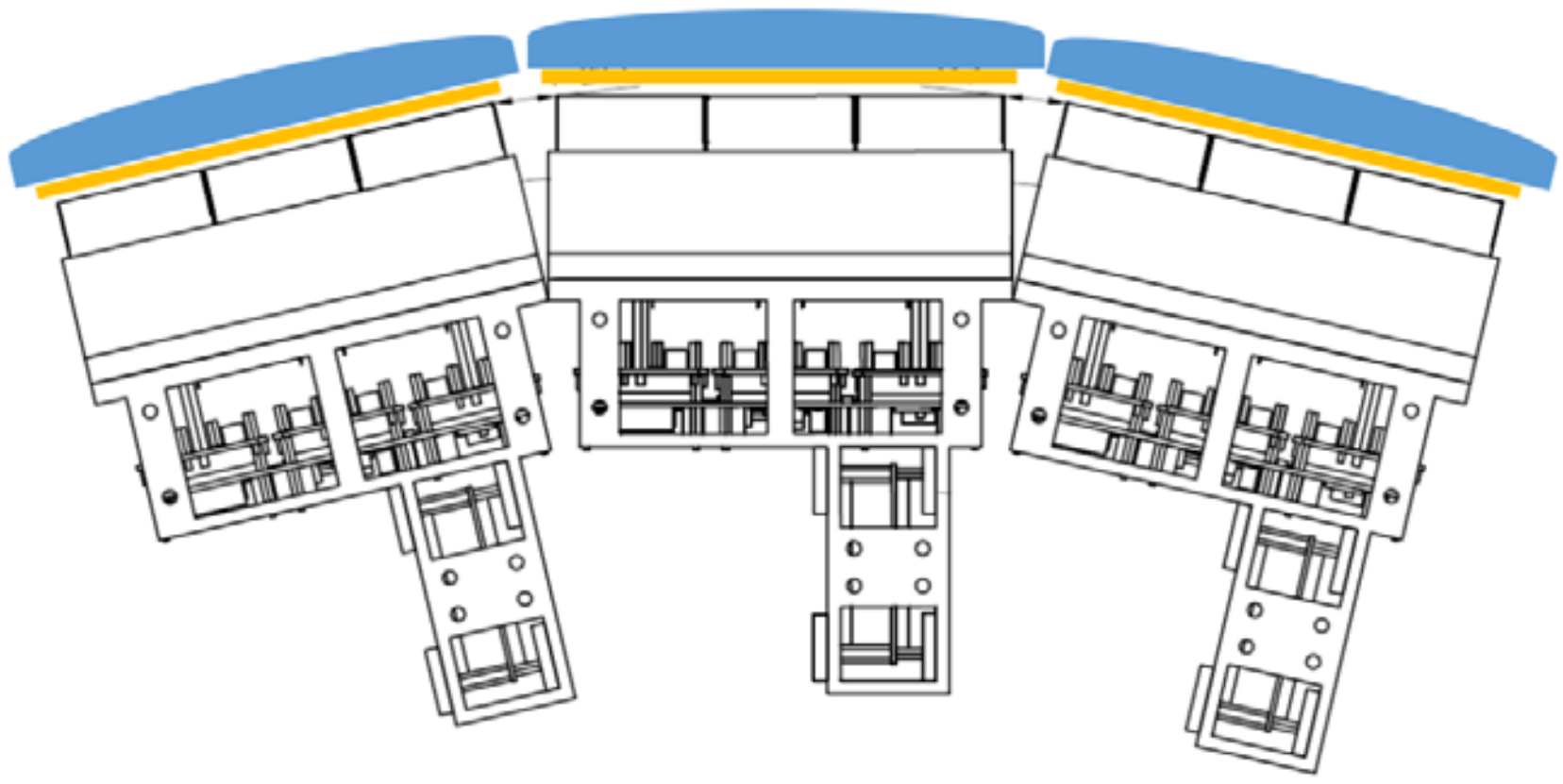

*Figure 3: Schematic view of the three PDMs covering the focal surface of the Schimdt optics.*

### 2.2.2 The UV camera electronics

The UV camera electronics is organized in three identical blocks, one for each PDM. The data from the 36 ASICs located on the same PDM are collected on three electronics boards (Cross boards) which perform the multiplexing of the signals. The multiplexed data are, in turn, sent to a second electronic board, called Zynq-board, containing a Xilinx FPGA with an embedded dual core ARM9 CPU processing system (Fig. 4 right panel). The Zynq board is responsible for the majority of the data handling, from data reception, buffering, configuration of the Spaciroc-3 ASICs and implementation of the trigger algorithms. This module also controls the HVPS board in order to have a real-time response to intense light signals as a second safety level against bright light. The Zynq-board stores the 1.0μs data stream in a running buffer on which runs the trigger code.

### 2.2.3 The trigger

The trigger algorithm implemented on the Zynq board should be able to recognize a fluorescence signal lasting a few tens of μs, while keeping the trigger rate on the level of 1 Hz/PDM. The general idea of the trigger logic is to have an adaptive threshold independent for each cluster of pixels, and then to count the number of active clusters in a certain portion of the PDM; an active cluster is defined as a cluster above its threshold.

The algorithm searches for a signal above n standard deviations from the average in any cluster of the focal surface. Both the rms and the average are calculated in real time to take into account varying illumination conditions. In case of a trigger, the 128 frame buffer (64 frames before the trigger and 64 after it) is stored in memory for each of the three PDMs. A detailed description of the trigger algorithm can be found in [9, 10].

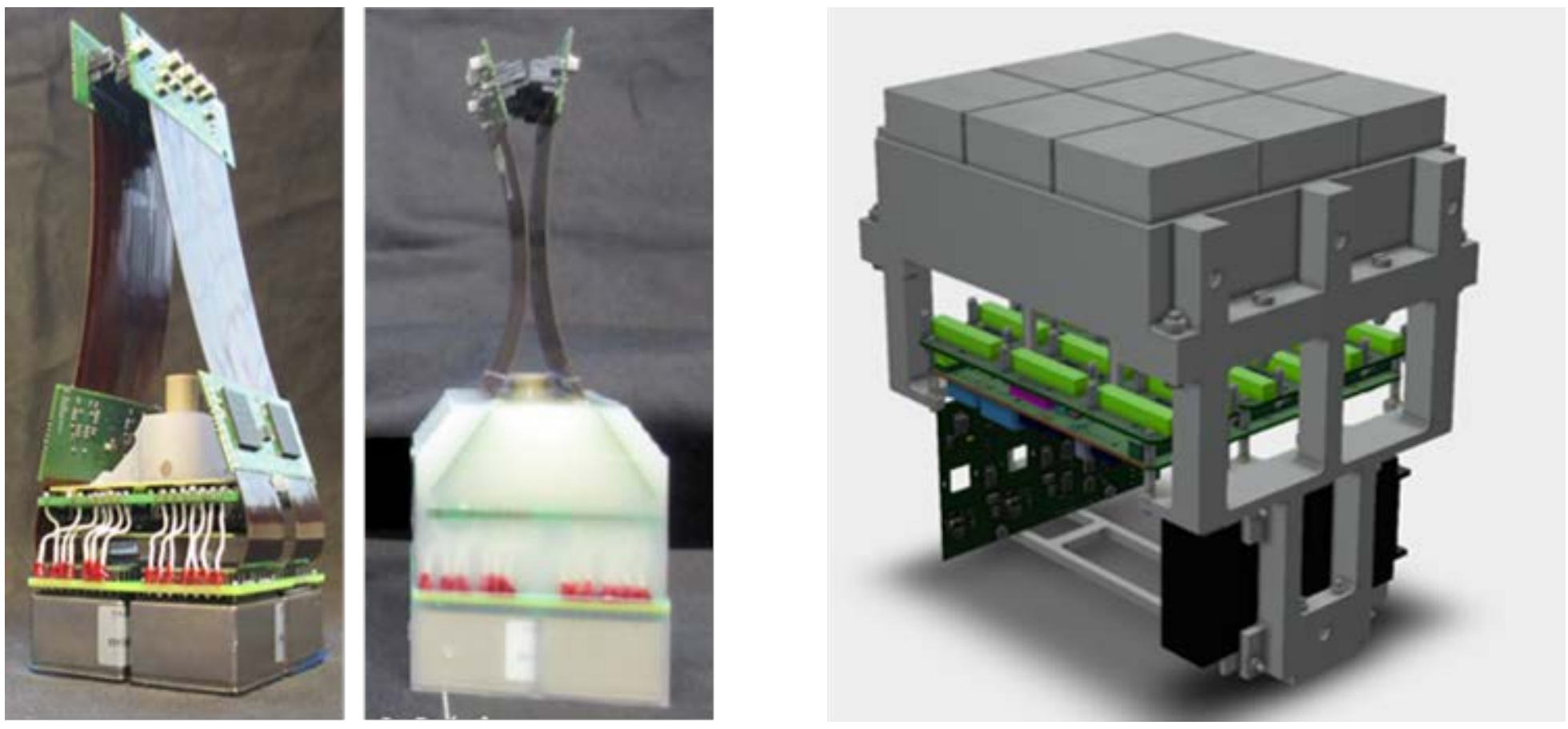

*Figure 4: Left: picture of the assembled EC before and after the "potting" procedure. Right: CAD rendering of the PDM. The UV camera electronics (Cross-boards and Zynq-board) is located on the back of the ECs.*

### 2.3 The Data Processor

The data processor (DP) is the primary interface with the telescope. It is a complex system that allows to remotely control, configure, monitor and operate the telescope, during integration phases, test and flight campaigns. The DP of the SPB2 Fluorescence telescope is an evolution of the one developed for EUSO-SPB1 [11-12]. The block diagram of the DP is shown in Fig.5. The DP is interfaced with the focal surface electronics through the Clock Board (CLKB). The CLKB board guarantees the time synchronization of the entire telescope and the tagging of each event acquired with the arrival time and the position of the balloon provided by the two GPS receivers.

The Central Processing Unit (CPU) module is the DP interface with the Gondola Control Computer (GCC) which communicates with NASA telemetry system. The data acquisition and the control of the UV camera electronics is performed via Ethernet connection. The data are saved on the 2 TB SATA raid array managed by the CPU. The limited bandwidth available on a super pressure flight constrains the percentage of the events that can be downloaded during the flight to a roughly 1% of the recorded events. Due to the relatively low probability to recover the payload at the end of the flight, machine learning techniques [10] will be used on on-board CPU to identify best UHECR candidate events and download them with the highest priority.

The Controller Area Network (CAN) bus is used to monitor the house-keeping parameters (voltages, currents and temperatures) of all the electronics systems. The CAN bus is also used to control and acquire the data of the auxiliary devices of the telescope (AMON, EMON, Health LED, etc.). The ability to perform orderly sequences of switching on and off of the subsystems is ensured by the presence of a Solid-State Power Module (SSPM). The SSPM is controlled by the CPU via the CAN bus as well.

The House Keeping (HK) system is responsible for acquiring the temperature sensors and controlling the heaters necessary to keep the telescope in the required temperature range. The designed architecture allows for a spare CPU on board that is able to take full control of the telescope in case of a main CPU failure.

The modules that constitute the DP are housed in two double-size Eurocard sub-racks.

### 2.4 The ancillary instruments

Airglow MONitor (AMON) and EUSO telescope darkness MONitor (EMON) detectors will be used as ancillary instruments. AMON [13] is a one-pixel instrument that will measure the absolute intensities of the radiation from the atmosphere below. Two AMON instruments will operate in the same wavelength bands of the FT to monitor airglow and night atmosphere radiation backgrounds. Two EMON instruments will be located inside telescope to monitor darkness conditions during operation and testing. Both the instruments use the CAN bus to transmit data and to receive configuration commands.

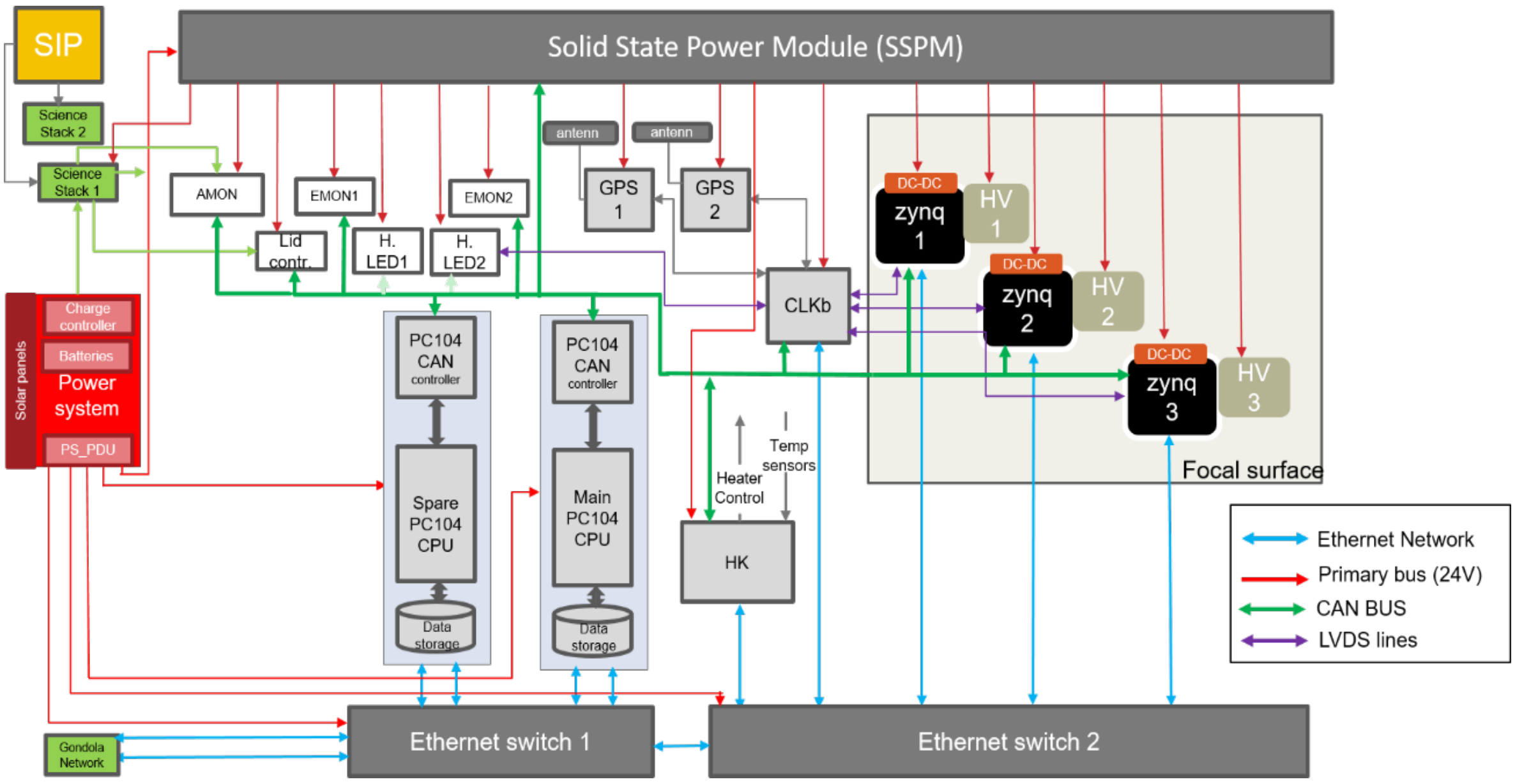

*Figure 5: The block diagram of the Data Processor system and FT sub-systems.*

The Health LED system provides a way to test the status of the UV camera pixels and, more in general, to test the health status of the whole telescope. It features a controller unit and a LED unit comprised of three LEDs. The Health LED system receives from the CLK board the signal to emit pulses at several frequencies or in correspondence of a GPS 1-PPS pulse.

## 3. Expected performance

Extensive simulation studies have been carried out to evaluate the expected events rate for the FT. The results show that the expected event rate is 0.12±0.01 UHECR events per hour of clear observation. The energy threshold is at $10^{18.2}$ eV, the peak energy sensitivity is at $10^{18.6}$ eV. A detailed description of the work is in [10].

## 4. Status and plan

All the components needed to build the telescope have been procured and the construction of its subsystems is now underway. Extensive laboratory and field testing is planned to characterize and calibrate the FT prior to payload integration. Its PSFs and optical throughput will be measured for downward and horizontal orientations using a 1 m diameter optical test beam [14]. The FT will then be transported to the Utah desert or for testing with lasers and other light sources. The whole instrument is on the way for a scheduled launch in early 2023 from Wanaka, New Zealand.

## Acknowledgements

This work was partially supported by Basic Science Interdisciplinary Research Projects of RIKEN and JSPS KAKENHI Grant (22340063, 23340081, and 24244042), by the Italian Ministry of Foreign Affairs and International Cooperation, by the Italian Space Agency through the ASI INFN agreements n. 2017-8-H.0 and n. 2021-8-HH.0, by NASA award 11-APRA-0058, 16-APROBES16-0023, 17-APRA17-0066, NNX17AJ82G, NNX13AH54G, 80NSSC18K0246, 80NSSC18K0473, 80NSSC19K0626, and 80NSSC18K0464 in the USA, by the French space agency CNES, by the Deutsches Zentrum für Luft- und Raumfahrt, the Helmholtz Alliance for Astroparticle Physics funded by the Initiative and Networking Fund of the Helmholtz Association (Germany), by Slovak Academy of Sciences MVTS JEM-EUSO, by National Science Centre in Poland grants 2017/27/B/ST9/02162 and 2020/37/B/ST9/01821, by Deutsche Forschungsgemeinschaft (DFG, German Research Foundation) under Germanys Excellence Strategy

- EXC-2094-390783311, by Mexican funding agencies PAPIIT-UNAM, CONACyT and the Mexican Space Agency (AEM), as well as VEGA grant agency project 2/0132/17, and by by State Space Corporation ROSCOSMOS and the Interdisciplinary Scientific and Educational School of Moscow University ”Fundamental and Applied Space Research”.

## References

[1] J.H. Adams, et al., *White paper on EUSO-SPB2*, 2017, arXiv:1703.04513v2 [astro-ph.HE].

[2] J. Eser et al., [JEM-EUSO Coll.], *Science and mission status of EUSO-SPB2*, in Proceedings, 37th International Cosmic Ray Conference (ICRC 2021): Berlin, Germany, July 12-23, 2021.

[3] A. Olinto, et al., *POEMMA: Probe of Extreme multi-messenger astrophysics*, 2017, ArXiv:170807599v1

[4] M. Bagheri et al., *Overview of Cherenkov Telescope onboard EUSO-SPB2 for the Detection of Ultra-High Energy Neutrinos*, in Proceedings, 37th International Cosmic Ray Conference (ICRC2021): Berlin, Germany, July 12-23, 2021.

[5] G. Abdellaoui, et al., JEM-EUSO Collaboration, *EUSO-TA–first results from a ground-based EUSO telescope, Astropart.* Phys. **102** (2018) 98–111.

[6] P. Von Ballmoos, et al., *A balloon-borne prototype for demonstrating the concept of JEM–EUSO*, Adv. Space Res. (ISSN: 02731177) **53** (10) (2014) 1544–1550,

[7] Osteria et al., *EUSO-SPB: In-flight performance*, Nuclear Instruments and Methods in Physics Research A **936** (2019) 237.

[8] J. Eser et al., [JEM-EUSO Coll.], *Results of the EUSO-SPB1 flight*, POS Proc. ICRC (2019).

[9] M. Bertaina et al., *The first level trigger of JEM-EUSO: Concept and tests*, Nuclear Instruments and Methods in Physics Research Section A: Accelerators, Spectrometers, Detectors and Associated Equipment **824** (2016) 253.

[10] G. Filippatos et al., [JEM-EUSO Coll.], *Expected Performance of the EUSO-SPB2 Fluorescence Telescope*, in Proceedings, 37th International Cosmic Ray Conference (ICRC 2021): Berlin, Germany, July 12-23,2021, 2021.

[11] V. Scotti, et al., *The data processor system of EUSO-SPB1*, Nucl. Instrum. Methods A **916** (2019) 94–101.

[12] C. Fornaro, et al., *The on-board software of the EUSO-SPB pathfinder experiment*, Softw. Pract. Exp. (2019)

[13] S. Mackovjak, et al., *Airglow monitor by one-pixel detector,* NIMA **922**, 150, (2019)

[14] V. Kungel et al., [JEM-EUSO Coll.], *EUSO-SPB2 Telescope Optics and Testing*, in Proceedings, 37th International Cosmic Ray Conference (ICRC 2021): Berlin, Germany, July 12-23,2021, 2021

# The JEM-EUSO Collaboration

(author-list as of July 1st, 2021)

G. Abdellaoui[ah], S. Abe[fq], J.H. Adams Jr.[pd], D. Allard[cb], G. Alonso[md], L. Anchordoqui[pe], A. Anzalone[eh,ed], E. Arnone[ek,el], K. Asano[fe], R. Attallah[ac], H. Attoui[aa], M. Ave Pernas[mc], M. Bagheri[ph], J. Baláz[la], M. Bakiri[aa], D. Barghini[el,ek], S. Bartocci[ei,ej], M. Battisti[ek,el], J. Bayer[dd], B. Beldjilali[ah], T. Belenguer[mb], N. Belkhalfa[aa], R. Bellotti[ea,eb], A.A. Belov[kb], K. Benmessai[aa], M. Bertaina[ek,el], P.F. Bertone[pf], P.L. Biermann[db], F. Bisconti[el,ek], C. Blaksley[ft], N. Blanc[oa], S. Blin-Bondil[ca,cb], P. Bobik[la], M. Bogomilov[ba], K. Bolmgren[na], E. Bozzo[ob], S. Briz[pb], A. Bruno[eh,ed], K.S. Caballero[hd], F. Cafagna[ea], G. Cambié[ei,ej], D. Campana[ef], J-N. Capdevielle[cb], F. Capel[de], A. Caramete[ja], L. Caramete[ja], P. Carlson[na], R. Caruso[ec,ed], M. Casolino[ft,ei], C. Cassardo[ek,el], A. Castellina[ek,em], O. Catalano[eh,ed], A. Cellino[ek,em], K. Černý[bb], M. Chikawa[fc], G. Chiritoi[ja], M.J. Christl[pf], R. Colalillo[ef,eg], L. Conti[en,ei], G. Cotto[ek,el], H.J. Crawford[pa], R. Cremonini[el], A. Creusot[cb], A. de Castro Gónzalez[pb], C. de la Taille[ca], L. del Peral[mc], A. Diaz Damian[cc], R. Diesing[pb], P. Dinaucourt[ca], A. Djakonow[ia], T. Djemil[ac], A. Ebersoldt[db], T. Ebisuzaki[ft], J. Eser[pb], F. Fenu[ek,el], S. Fernández-González[ma], S. Ferrarese[ek,el], G. Filippatos[pc], W.I. Finch[pc] C. Fornaro[en,ei], M. Fouka[ab], A. Franceschi[ee], S. Franchini[md], C. Fuglesang[na], T. Fujii[fg], M. Fukushima[fe], P. Galeotti[ek,el], E. García-Ortega[ma], D. Gardiol[ek,em], G.K. Garipov[kb], E. Gascón[ma], E. Gazda[ph], J. Genci[lb], A. Golzio[ek,el], C. González Alvarado[mb], P. Gorodetzky[ft], A. Green[pc], F. Guarino[ef,eg], C. Guépin[pl], A. Guzmán[dd], Y. Hachisu[ft], A. Haungs[db], J. Hernández Carretero[mc], L. Hulett[pc], D. Ikeda[fe], N. Inoue[fn], S. Inoue[ft], F. Isgrò[ef,eg], Y. Itow[fk], T. Jammer[dc], S. Jeong[gb], E. Joven[me], E.G. Judd[pa], J. Jochum[dc], F. Kajino[ff], T. Kajino[fi], S. Kalli[af], I. Kaneko[ft], Y. Karadzhov[ba], M. Kasztelan[ia], J. Katahira[ft], K. Kawai[ft], Y. Kawasaki[ft], A. Kedadra[aa], H. Khales[aa], B.A. Khrenov[kb], Jeong-Sook Kim[ga], Soon-Wook Kim[ga], M. Kleifges[db], P.A. Klimov[kb], D. Kolev[ba], I. Kreykenbohm[da], J.F. Krizmanic[pf,pk], K. Królik[ia], V. Kungel[pc], Y. Kurihara[fs], A. Kusenko[fr,pe], E. Kuznetsov[pd], H. Lahmar[aa], F. Lakhdari[ag], J. Licandro[me], L. López Campano[ma], F. López Martínez[pb], S. Mackovjak[la], M. Mahdi[aa], D. Mandát[bc], M. Manfrin[ek,el], L. Marcelli[ei], J.L. Marcos[ma], W. Marszał[ia], Y. Martín[me], O. Martinez[hc], K. Mase[fa], R. Matev[ba], J.N. Matthews[pg], N. Mebarki[ad], G. Medina-Tanco[ha], A. Menshikov[db], A. Merino[ma], M. Mese[ef,eg], J. Meseguer[md], S.S. Meyer[pb], J. Mimouni[ad], H. Miyamoto[ek,el], Y. Mizumoto[fi], A. Monaco[ea,eb], J.A. Morales de los Ríos[mc], M. Mastafa[pd], S. Nagataki[ft], S. Naitamor[ab], T. Napolitano[ee], J. M. Nachtman[pi] A. Neronov[ob,cb], K. Nomoto[fr], TNonaka[fe], T. Ogawa[ft], S. Ogio[fl], H. Ohmori[ft], A.V. Olinto[pb], Y. Onel[pi] G. Osteria[ef], A.N. Otte[ph], A. Pagliaro[eh,ed], W. Painter[db], M.I. Panasyuk[kb], B. Panico[ef], E. Parizot[cb], I.H. Park[gb], B. Pastircak[la], T. Paul[pe], M. Pech[bb], I. Pérez-Grande[md], F. Perfetto[ef], T. Peter[oc], P. Picozza[ei,ej,ft], S. Pindado[md], L.W. Piotrowski[ib], S. Piraino[dd], Z. Plebaniak[ek,el,ia], A. Pollini[oa], E.M. Popescu[ja], R. Prevete[ef,eg], G. Prévôt[cb], H. Prieto[mc], M. Przybylak[ia], G. Puehlhofer[dd], M. Putis[la], P. Reardon[pd], M.H.. Reno[pi], M. Reyes[me], M. Ricci[ee], M.D. Rodríguez Frías[mc], O.F. Romero Matamala[ph], F. Ronga[ee], M.D. Sabau[mb], G. Saccá[ec,ed], G. Sáez Cano[mc], H. Sagawa[fe], Z. Sahnoune[ab], A. Saito[fg], N. Sakaki[ft], H. Salazar[hc], J.C. Sanchez Balanzar[ha], J.L. Sánchez[ma], A. Santangelo[dd], A. Sanz-Andrés[md], M. Sanz Palomino[mb], O.A. Saprykin[kc], F. Sarazin[pc], M. Sato[fo], A. Scagliola[ea,eb], T. Schanz[dd], H. Schieler[db], P. Schovánek[bc], V. Scotti[ef,eg], M. Serra[me], S.A. Sharakin[kb], H.M. Shimizu[fj], K. Shinozaki[ia], J.F. Soriano[pe], A. Sotgiu[ei,ej], I. Stan[ja], I. Strharský[la], N. Sugiyama[fj], D. Supanitsky[ha], M. Suzuki[fm], J. Szabelski[ia], N. Tajima[ft], T. Tajima[ft], Y. Takahashi[fo], M. Takeda[fe], Y. Takizawa[ft], M.C. Talai[ac], Y. Tameda[fp], C. Tenzer[dd], S.B. Thomas[pg], O. Tibolla[he], L.G. Tkachev[ka], T. Tomida[fh], N. Tone[ft], S. Toscano[ob], M. Traïche[aa], Y. Tsunesada[fl], K. Tsuno[ft], S. Turriziani[ft], Y. Uchihori[fb], O. Vaduvescu[me], J.F. Valdés-Galicia[ha], P. Vallania[ek,em], L. Valore[ef,eg], G. Vankova-Kirilova[ba], T. M. Venters[pj], C. Vigorito[ek,el], L. Villaseñor[hb], B. Vlcek[mc], P. von Ballmoos[cc], M. Vrabel[lb], S. Wada[ft], J. Watanabe[fi], J. Watts Jr.[pd], R. Weigand Muñoz[ma], A. Weindl[db], L. Wiencke[pc], M. Wille[da], J. Wilms[da], D. Winn[pm] T. Yamamoto[ff], J. Yang[gb], H. Yano[fm], I.V. Yashin[kb], D. Yonetoku[fd], S. Yoshida[fa], R. Young[pf], I.S Zgura[ja],

M.Yu. Zotov[kb], A. Zuccaro Marchi[ft]

[aa] Centre for Development of Advanced Technologies (CDTA), Algiers, Algeria
[ab] Dep. Astronomy, Centre Res. Astronomy, Astrophysics and Geophysics (CRAAG), Algiers, Algeria
[ac] LPR at Dept. of Physics, Faculty of Sciences, University Badji Mokhtar, Annaba, Algeria
[ad] Lab. of Math. and Sub-Atomic Phys. (LPMPS), Univ. Constantine I, Constantine, Algeria
[af] Department of Physics, Faculty of Sciences, University of M'sila, M'sila, Algeria
[ag] Research Unit on Optics and Photonics, UROP-CDTA, Sétif, Algeria
[ah] Telecom Lab., Faculty of Technology, University Abou Bekr Belkaid, Tlemcen, Algeria
[ba] St. Kliment Ohridski University of Sofia, Bulgaria
[bb] Joint Laboratory of Optics, Faculty of Science, Palacký University, Olomouc, Czech Republic
[bc] Institute of Physics of the Czech Academy of Sciences, Prague, Czech Republic
[ca] Omega, Ecole Polytechnique, CNRS/IN2P3, Palaiseau, France
[cb] Université de Paris, CNRS, AstroParticule et Cosmologie, F-75013 Paris, France
[cc] IRAP, Université de Toulouse, CNRS, Toulouse, France
[da] ECAP, University of Erlangen-Nuremberg, Germany
[db] Karlsruhe Institute of Technology (KIT), Germany
[dc] Experimental Physics Institute, Kepler Center, University of Tübingen, Germany
[dd] Institute for Astronomy and Astrophysics, Kepler Center, University of Tübingen, Germany
[de] Technical University of Munich, Munich, Germany
[ea] Istituto Nazionale di Fisica Nucleare - Sezione di Bari, Italy
[eb] Universita' degli Studi di Bari Aldo Moro and INFN - Sezione di Bari, Italy
[ec] Dipartimento di Fisica e Astronomia "Ettore Majorana", Universita' di Catania, Italy
[ed] Istituto Nazionale di Fisica Nucleare - Sezione di Catania, Italy
[ee] Istituto Nazionale di Fisica Nucleare - Laboratori Nazionali di Frascati, Italy
[ef] Istituto Nazionale di Fisica Nucleare - Sezione di Napoli, Italy
[eg] Universita' di Napoli Federico II - Dipartimento di Fisica "Ettore Pancini", Italy
[eh] INAF - Istituto di Astrofisica Spaziale e Fisica Cosmica di Palermo, Italy
[ei] Istituto Nazionale di Fisica Nucleare - Sezione di Roma Tor Vergata, Italy
[ej] Universita' di Roma Tor Vergata - Dipartimento di Fisica, Roma, Italy
[ek] Istituto Nazionale di Fisica Nucleare - Sezione di Torino, Italy
[el] Dipartimento di Fisica, Universita' di Torino, Italy
[em] Osservatorio Astrofisico di Torino, Istituto Nazionale di Astrofisica, Italy
[en] Uninettuno University, Rome, Italy
[fa] Chiba University, Chiba, Japan
[fb] National Institutes for Quantum and Radiological Science and Technology (QST), Chiba, Japan
[fc] Kindai University, Higashi-Osaka, Japan
[fd] Kanazawa University, Kanazawa, Japan
[fe] Institute for Cosmic Ray Research, University of Tokyo, Kashiwa, Japan
[ff] Konan University, Kobe, Japan
[fg] Kyoto University, Kyoto, Japan
[fh] Shinshu University, Nagano, Japan
[fi] National Astronomical Observatory, Mitaka, Japan
[fj] Nagoya University, Nagoya, Japan
[fk] Institute for Space-Earth Environmental Research, Nagoya University, Nagoya, Japan
[fl] Graduate School of Science, Osaka City University, Japan
[fm] Institute of Space and Astronautical Science/JAXA, Sagamihara, Japan
[fn] Saitama University, Saitama, Japan
[fo] Hokkaido University, Sapporo, Japan

[fp] Osaka Electro-Communication University, Neyagawa, Japan
[fq] Nihon University Chiyoda, Tokyo, Japan
[fr] University of Tokyo, Tokyo, Japan
[fs] High Energy Accelerator Research Organization (KEK), Tsukuba, Japan
[ft] RIKEN, Wako, Japan
[ga] Korea Astronomy and Space Science Institute (KASI), Daejeon, Republic of Korea
[gb] Sungkyunkwan University, Seoul, Republic of Korea
[ha] Universidad Nacional Autónoma de México (UNAM), Mexico
[hb] Universidad Michoacana de San Nicolas de Hidalgo (UMSNH), Morelia, Mexico
[hc] Benemérita Universidad Autónoma de Puebla (BUAP), Mexico [hd] Universidad Autónoma de Chiapas (UNACH), Chiapas, Mexico [he] Centro Mesoamericano de Física Teórica (MCTP), Mexico
[ia] National Centre for Nuclear Research, Lodz, Poland
[ib] Faculty of Physics, University of Warsaw, Poland [ja] Institute of Space Science ISS, Magurele, Romania [ka] Joint Institute for Nuclear Research, Dubna, Russia
[kb] Skobeltsyn Institute of Nuclear Physics, Lomonosov Moscow State University, Russia
[kc] Space Regatta Consortium, Korolev, Russia
[la] Institute of Experimental Physics, Kosice, Slovakia
[lb] Technical University Kosice (TUKE), Kosice, Slovakia
[ma] Universidad de León (ULE), León, Spain
[mb] Instituto Nacional de Técnica Aeroespacial (INTA), Madrid, Spain
[mc] Universidad de Alcalá (UAH), Madrid, Spain
[md] Universidad Politécnia de madrid (UPM), Madrid, Spain [me] Instituto de Astrofísica de Canarias (IAC), Tenerife, Spain [na] KTH Royal Institute of Technology, Stockholm, Sweden
[oa] Swiss Center for Electronics and Microtechnology (CSEM), Neuchâtel, Switzerland
[ob] ISDC Data Centre for Astrophysics, Versoix, Switzerland
[oc] Institute for Atmospheric and Climate Science, ETH Zürich, Switzerland [pa] Space Science Laboratory, University of California, Berkeley, CA, USA [pb] University of Chicago, IL, USA
[pc] Colorado School of Mines, Golden, CO, USA
[pd] University of Alabama in Huntsville, Huntsville, AL; USA
[pe] Lehman College, City University of New York (CUNY), NY, USA
[pf] NASA Marshall Space Flight Center, Huntsville, AL, USA
[pg] University of Utah, Salt Lake City, UT, USA
[ph] Georgia Institute of Technology, USA
[pi] University of Iowa, Iowa City, IA, USA
[pj] NASA Goddard Space Flight Center, Greenbelt, MD, USA
[pk] Center for Space Science & Technology, University of Maryland, Baltimore County, Baltimore, MD, USA
[pl] Department of Astronomy, University of Maryland, College Park, MD, USA
[pm] Fairfield University, Fairfield, CT, USA